# Power System Differential-Algebraic Equations

Bin Wang, Yang Liu and Kai Sun

***Abstract*—This document presents an introduction of two commonly used power system differential algebraic equations for studying electromechanical oscillation and transient stability. Two types of generator models are used to formulate the power system model, respectively: the second-order classical model and the fourth-order generator model. An example is provided on the IEEE 9-bus system.**

## I. INTRODUCTION

POWER systems are nonlinear dynamic systems, whose behaviors are usually modeled by differential-algebraic equations (DAEs). The algebraic equations describe the network connectivity and its parameters, coordinate transformations used for generator equations, and some static elements, e.g. static loads. The differential equations describe the behaviors of all dynamic elements, including generators and their control systems, dynamic loads and FACTs (flexible alternating current transmission systems) devices.

The study of a general power system model might be difficult and unnecessary, since the time constants for different dynamic elements are quite different, ranging from $10^{-4}$s, e.g. the switching time constant of power electronic devices like FACTs, to $10^{1}$s, e.g. the governor system of generators. Thus, for a specific study purpose, a comment handling is to consider the fast and slow dynamics separately, which means only part of the equations corresponding to the elements with interesting behaviors will be remained. In particular, for studying the electromechanical oscillation and transient stability a power system, equations associated with generator, exciter and governor models are usually remained. For simplicity, this document will only cover the generator model and point out where to add the state variables when considering the exciter and governor models, and all load are in constant impedance model (a static model).

Two basic state variables of a generator are the rotor angle $\delta$ and the rotor speed $\omega$. For any stable power system, the rotor angles of all generators are close to each other but constantly changing at a certain synchronized speed/frequency which is slightly floating around 60Hz/50Hz. Since only the relative motions among different angles are of concern when analyzing the oscillation or stability, a rotating coordinate system at the synchronized speed/frequency is commonly used to represent the system equations, where the rotor speed $\Delta\omega$ is zero and rotor angle $\Delta\delta$ is a constant for each generator at the system equilibrium. The rest of the document is based on the rotating coordinate system.

The generator can be modeled to different detailed levels for different purposes of study. This document introduces the power system DAEs with either the simplest generator model, i.e. second-order classical model, or the fourth-order model.

B. Wang, Y. Liu and K. Sun are with the University of Tennessee, Knoxville, TN 37996 USA (e-mails: bwang13@vols.utk.edu, yliu161@vols.utk.edu, kaisun@utk.edu).

## II. POWER SYSTEM DIFFERENTIAL-ALGEBRAIC EQUATIONS

### A. With classical generator model

The classical generator model is a second-order differential equation with two state variables: $\delta$ and $\Delta\omega$. The DAEs of a power system with $m$-generator are:

$$\begin{cases} \dot{\delta}_i = \omega_s \Delta\omega_i \\ \Delta\dot{\omega}_i = \dfrac{1}{2H_i}\left(P_{mi} - P_{ei} - D_i \Delta\omega_i\right) \end{cases} \tag{1}$$

where $\delta_i$ and $\Delta\omega_i$ are the rotor angle and rotor speed derivation from the nominal value of generator $i$, respectively; $P_{ei}$ is the electric power of generator $i$, which is a function of all rotor angles and network parameters; $H_i$ and $D_i$ are inertia and damping constants of generator $i$, respectively; $P_{mi}$ is the mechanical power of generator $i$; $\omega_s$ is the synchronized frequency of the system.

The expression of $P_{ei}$ is:

$$\begin{cases} \begin{bmatrix} e'_{xi} \\ e'_{yi} \end{bmatrix} = \begin{bmatrix} \sin\delta_i & \cos\delta_i \\ -\cos\delta_i & \sin\delta_i \end{bmatrix} \begin{bmatrix} 0 \\ e'_{qi} \end{bmatrix} \\ I_t = \left[ i_{x1}, i_{y1}, \cdots, i_{xm}, i_{ym} \right]^T \\ \quad = Y\left[ e'_{x1}, e'_{y1}, \cdots, e'_{xm},, e'_{ym} \right]^T \\ \begin{bmatrix} e_{xi} \\ e_{yi} \end{bmatrix} = \begin{bmatrix} e'_{qi}\cos\delta_i \\ e'_{qi}\sin\delta_i \end{bmatrix} - \begin{bmatrix} R_{ai} & -X'_{di} \\ X'_{di} & R_{ai} \end{bmatrix} \begin{bmatrix} i_{xi} \\ i_{yi} \end{bmatrix} \\ P_{ei} = e_{xi} i_{xi} + e_{yi} i_{yi} \end{cases} \tag{2}$$

where $e'_{qi}$ is the field voltage of generator $i$ which can be calculated from the initial condition of the system; $e'_{xi}$ and $e'_{yi}$ are the internal bus voltages of generator $i$ in the non-rotating coordinate; $Y_i$ is the $i$th row of the admittance matrix of the reduced network, see Appendix A, including source impedances of generators and constant-impedance loads; $I_{ti}$ is the terminal current of generator $i$; $R_{ai}$ and $X'_{di}$ are the source impedance of generator $i$; $e_{xi}$ and $e_{xi}$ are terminal voltages for $x$ and $y$ axes of generator $i$. Note that $Y$ matrix is a constant matrix, see proof in Appendix C.

### B. With fourth-order generator model

The forth-order generator model is a fourth-order differential equation with four state variables: $\delta$, $\Delta\omega$, $e'_q$ and $e'_d$. The DAEs of a power system with $m$-generator are [1]:

$$\begin{cases} \dot{\delta}_i = \omega_s \Delta\omega_i \\ \Delta\dot{\omega}_i = \dfrac{1}{2H_i}\left(P_{mi} - P_{ei} - D_i \Delta\omega_i\right) \\ \dot{e}'_{qi} = \dfrac{1}{T'_{d0i}}\left[E_{fdi} - (X_{di} - X'_{di}) i_{di} - e'_{qi}\right] \\ \dot{e}'_{di} = \dfrac{1}{T'_{q0i}}\left[(X_{qi} - X'_{qi}) i_{qi} - e'_{di}\right] \end{cases} \tag{3}$$

where $\delta_i$, $\Delta\omega_i$, $e'_{qi}$ and $e'_{di}$ are the rotor angle, rotor speed derivation from the nominal value, transient voltages along $q$ and $d$ axes respectively of generator $i$; $P_{ei}$, $i_{di}$ and $i_{qi}$ are the electric power, stator currents of $q$ and $d$ axes respectively of generator $i$, which are functions of all rotor angles and network parameters; $H_i$ and $D_i$ are inertia and damping constants of generator $i$; $T'_{d0}$ and $T'_{q0}$ are the open-circuit time constants, $X_d$ and $X_q$ are the synchronous reactance, $X'_d$ and $X'_q$ are the transient reactance respectively for $q$ and $d$ axes of generator $i$; $P_{mi}$ is the mechanical power of generator $i$; $\omega_s$ is the synchronized frequency of the system.

The expressions of $P_{ei}$, $i_{di}$ and $i_{qi}$ are:

$$\begin{cases} \begin{bmatrix} e'_{xi} \\ e'_{yi} \end{bmatrix} = \begin{bmatrix} \sin\delta_i & \cos\delta_i \\ -\cos\delta_i & \sin\delta_i \end{bmatrix} \begin{bmatrix} e'_{di} \\ e'_{qi} \end{bmatrix} \\ I_t = \left[i_{x1}, i_{y1}, \cdots, i_{xm}, i_{ym}\right]^T \\ \quad = Y\left[e'_{x1}, e'_{y1}, \cdots, e'_{xm},, e'_{ym}\right] \\ \begin{bmatrix} i_{di} \\ i_{qi} \end{bmatrix} = \begin{bmatrix} \sin\delta_i & -\cos\delta_i \\ \cos\delta_i & \sin\delta_i \end{bmatrix} \begin{bmatrix} i_{xi} \\ i_{yi} \end{bmatrix} \\ \begin{bmatrix} e_{di} \\ e_{qi} \end{bmatrix} = \begin{bmatrix} e'_{di} \\ e'_{qi} \end{bmatrix} - \begin{bmatrix} R_{ai} & -X'_{qi} \\ X'_{di} & R_{ai} \end{bmatrix} \begin{bmatrix} i_{di} \\ i_{qi} \end{bmatrix} \\ P_{ei} = e_{di} i_{di} + e_{qi} i_{qi} \end{cases} \tag{4}$$

where $e'_{xi}$ and $e'_{yi}$ are the transient voltages in the non-rotating coordinate representation of generator $i$; $Y_i$ is the $i$th row of the admittance matrix of the reduced network, see Appendix A; $I_{ti}$ is the terminal current of generator $i$; $R_{ai}$ is the resistance of generator $i$; $i_{qi}$ and $i_{di}$ are terminal current, $e_{qi}$ and $e_{di}$ are terminal voltages for $q$ and $d$ axes of generator $i$. Note that the general $Y$ depends on all rotor angles and should be updated during each integration step, while $Y$ will become a constant matrix when $X'_{di}$=$X'_{qi}$ holds for all generators, see proof in Appendix C.

### C. Consideration of exciter and governor models

When adding an exciter model to the generator $i$, the parameter $E_{fdi}$ becomes a state variable dominated by the differential equation(s) associated with the exciter, while all other equations of generator $i$ remain the same. Similarly, the parameter $P_{mi}$ will become a differential variable when adding a governor model to generator $i$.

It can be seen from (1) that the classical generator model accepts a governor model, but it cannot accept an exciter model since its field voltage is modeled as a constant and cannot be a state variable of some differential equation. The fourth-order generator model in (3) can accept both of exciter and governor models.

For the generator without exciter and governor, its $P_m$ and $E_{fd}$ are constants and determined by the initial conditions.

## III. Example on the IEEE 9-Bus Power System

This section presents the DAEs of the IEEE 3-machine, 9-bus power system. Since the equations are shown through (1) to (4), this section will only provide an example of all parameters and initial conditions based on [2]. Note that the synchronized speed/frequency usually takes $2\pi f_s$, where $f_s$=60Hz or 50Hz, and 60Hz is used in this document.

### A. With classical generator model

Table I. Parameters and initial values of the variables for system with the classical generator model

| | Generator 1 | Generator 2 | Generator 3 |
|---|---|---|---|
| $H$ | 23.64 | 6.40 | 3.01 |
| $D$ | 23.64 | 6.40 | 3.01 |
| $R_a$ | 0 | 0 | 0 |
| $X'_d$ | 0.0608 | 0.1198 | 0.1813 |
| $P_m$ | 0.7164 | 1.6300 | 0.8500 |
| $e'_q$ | 1.0566 | 1.0502 | 1.0170 |
| $\delta(0)$ | 0.0626 | 1.0567 | 0.9449 |
| $\Delta\omega(0)$ | 0 | 0 | 0 |
| $Real(I_t)$ | 0.6889 | 1.5799 | 0.8179 |
| $Imag(I_t)$ | -0.2601 | 0.1924 | 0.1730 |
| $I_d$ | 0.2872 | 0.3523 | 0.0178 |
| $I_q$ | 0.6780 | 1.5521 | 0.8358 |

Parameters: $H$, $D$, $R_a$, $X'_d$

Constants: $P_m$, $e'_q$

Initial values of differential variables: $\delta$, $\Delta\omega$

Initial values of algebraic variables: $I_t$, $I_d$, $I_q$

This document does not have an introduction on how to calculate those initial values of differential and algebraic variables. For those who are interested in this, please refer to some power system analysis textbook, e.g. [1].

For the system in the steady-state with no-disturbance, the admittance $Y$, reduced admittance matrix $Y_t$ and the matrix $T_2$ are shown below.

$$Y = \begin{bmatrix} 0.8455 & 2.9883 & 0.2871 & -1.5129 & 0.2096 & -1.2256 \\ -2.9883 & 0.8455 & 1.5129 & 0.2871 & 1.2256 & 0.2096 \\ 0.2871 & -1.5129 & 0.4200 & 2.7239 & 0.2133 & -1.0879 \\ 1.5129 & 0.2871 & -2.7239 & 0.4200 & 1.0879 & 0.2133 \\ 0.2096 & -1.2256 & 0.2133 & -1.0879 & 0.2770 & 2.3681 \\ 1.2256 & 0.2096 & 1.0879 & 0.2133 & -2.3681 & 0.2770 \end{bmatrix} \tag{5}$$

$$Y_t = \begin{bmatrix} 1.1051 - j4.6957 & 0.0965 + j2.2570 & 0.0046 + j2.2748 \\ 0.0965 + j2.2570 & 0.7355 - j5.1143 & 0.1230 + j2.8257 \\ 0.0046 + j2.2748 & 0.1230 + j2.8257 & 0.7214 - j5.0231 \end{bmatrix} \quad (6)$$

$$T_2 = \begin{bmatrix} 0 & -0.0608 & 0 & 0 & 0 & 0 \\ 0.0608 & 0 & 0 & 0 & 0 & 0 \\ 0 & 0 & 0 & -0.1198 & 0 & 0 \\ 0 & 0 & 0.1198 & 0 & 0 & 0 \\ 0 & 0 & 0 & 0 & 0 & -0.1813 \\ 0 & 0 & 0 & 0 & 0.1813 & 0 \end{bmatrix} \quad (7)$$

An example of $Y$ matrices for fault-on and post-fault system under a specific fault are provided in the appendix D [2].

### B. *With fourth-order generator model*

Table II. Parameters and initial values of the variables with the fourth-order generator model

| | Generator 1 | Generator 2 | Generator 3 |
|---|---|---|---|
| $H$ | 23.64 | 6.40 | 3.01 |
| $D$ | 0 | 0 | 0 |
| $X_d$ | 0.1460 | 0.8958 | 1.3125 |
| $X_q$ | 0.0969 | 0.8465 | 1.2578 |
| $X'_d$ | 0.0608 | 0.1198 | 0.1813 |
| $X'_q$ | 0.0969 | 0.1969 | 0.2500 |
| $T'_{d0}$ | 8.9600 | 6.0000 | 5.8900 |
| $T'_{q0}$ | 0.6000 | 0.5350 | 0.6000 |
| $R_a$ | 0 | 0 | 0 |
| $\delta(0)$ | 0.0626 | 1.0664 | 0.9449 |
| $\Delta\omega(0)$ | 0 | 0 | 0 |
| $e'_q(0)$ | 1.0564 | 0.7882 | 0.7679 |
| $e'_d(0)$ | 0 | 0.6222 | 0.6242 |
| $P_m$ | 0.7164 | 1.6300 | 0.8500 |
| $E_{fd}$ | 1.0821 | 1.7893 | 1.4030 |
| $Real(I_t)$ | 0.6889 | 1.5799 | 0.8179 |
| $Imag(I_t)$ | -0.2600 | 0.1924 | 0.1730 |
| $I_d$ | 0.3026 | 1.2901 | 0.5615 |
| $I_q$ | 0.6712 | 0.9320 | 0.6194 |

Parameters: $H$, $D$, $X_d$, $X_q$, $X'_d$, $X'_q$, $T'_{d0}$, $T'_{q0}$, $R_a$

Constants: $P_m$, $E_{fd}$

Initial values of differential variables: $\delta$, $\Delta\omega$, $e'_q$, $e'_d$

Initial values of algebraic variables: $I_t$, $I_d$, $I_q$

The admittance matrix $Y_t$ is the same as the one in (5) and the matrix $T_2$ is shown below. Note that the matrix $Y$ depends on all rotor angles, which is not a constant matrix.

$$T_2 = \begin{bmatrix} 0 & -0.0969 & 0 & 0 & 0 & 0 \\ 0.0608 & 0 & 0 & 0 & 0 & 0 \\ 0 & 0 & 0 & -0.1969 & 0 & 0 \\ 0 & 0 & 0.1198 & 0 & 0 & 0 \\ 0 & 0 & 0 & 0 & 0 & -0.2500 \\ 0 & 0 & 0 & 0 & 0.1813 & 0 \end{bmatrix} \quad (8)$$

Consider another system similar to the one in Table II with the only difference that $X'_q=X'_d$ for each generator, the matrix $Y$ will become a constant. The following shows the different parameters and values compared to those in Table II where $X'_q \neq X'_d$.

Table III. Parameters and initial values that are different from the system in Table II

| | Generator 1 | Generator 2 | Generator 3 |
|---|---|---|---|
| $X'_q$ | 0.0608 | 0.1198 | 0.1813 |
| $e'_d(0)$ | 0.0242 | 0.6941 | 0.6668 |

$$Y = \begin{bmatrix} 0.8455 & 2.9883 & 0.2871 & -1.5129 & 0.2096 & -1.2256 \\ -2.9883 & 0.8455 & 1.5129 & 0.2871 & 1.2256 & 0.2096 \\ 0.2871 & -1.5129 & 0.4200 & 2.7239 & 0.2133 & -1.0879 \\ 1.5129 & 0.2871 & -2.7239 & 0.4200 & 1.0879 & 0.2133 \\ 0.2096 & -1.2256 & 0.2133 & -1.0879 & 0.2770 & 2.3681 \\ 1.2256 & 0.2096 & 1.0879 & 0.2133 & -2.3681 & 0.2770 \end{bmatrix} \quad (9)$$

$$T_2 = \begin{bmatrix} 0 & -0.0608 & 0 & 0 & 0 & 0 \\ 0.0608 & 0 & 0 & 0 & 0 & 0 \\ 0 & 0 & 0 & -0.1198 & 0 & 0 \\ 0 & 0 & 0.1198 & 0 & 0 & 0 \\ 0 & 0 & 0 & 0 & 0 & -0.1813 \\ 0 & 0 & 0 & 0 & 0.1813 & 0 \end{bmatrix} \quad (10)$$

## V. Appendix A

This appendix shows how to calculate $Y$, i.e. the admittance matrix of the reduced network to internal buses of all generators, from $Y_t$, whose calculation is introduced in Appendix B. In general, matrix $Y$ is not constant. In some special cases, $Y$ will become a constant matrix, which will be introduced and proved in Appendix C.

The matrix $Y$ is defined in

$$Y = \left(T_1 Y_r^{-1} + T_2 T_1\right)^{-1} T_1 \quad (A1)$$

where $Y_r$ is the admittance matrix reduced to the terminal bus of every generator, see its calculation in Appendix B. Note that $Y_r$ is a $(2m)\times(2m)$ matrix.

$$T_1 = \begin{pmatrix} \Delta_1 & 0 & 0 & 0 \\ 0 & \Delta_2 & 0 & 0 \\ 0 & 0 & \ddots & 0 \\ 0 & 0 & 0 & \Delta_m \end{pmatrix}, \quad T_2 = \begin{pmatrix} Z_1 & 0 & 0 & 0 \\ 0 & Z_2 & 0 & 0 \\ 0 & 0 & \ddots & 0 \\ 0 & 0 & 0 & Z_m \end{pmatrix} \quad (A2)$$

$$\Delta_k = \begin{pmatrix} \sin\delta_k & -\cos\delta_k \\ \cos\delta_k & \sin\delta_k \end{pmatrix} \quad (A3)$$

$$Z_k = \begin{pmatrix} R_{ak} & -X'_{qk} \\ X'_{dk} & R_{ak} \end{pmatrix} \qquad (A4)$$

## VI. APPENDIX B

This appendix shows how to calculate $Y_r$, the admittance matrix of the reduced network to terminal buses of all generators, from $Y_{bus}$, the bus admittance matrix.

Node voltage equation with ground as reference gives:

$$I_{bus} = Y_{bus} V_{bus}$$

$$\begin{bmatrix} I_1 \\ I_2 \\ \vdots \\ I_n \\ \hline I_{n+1} \\ \vdots \\ I_{n+m} \end{bmatrix} = \left[\begin{array}{ccc|ccc} Y_{11} & \cdots & Y_{1n} & Y_{1(n+1)} & \cdots & Y_{1(n+m)} \\ Y_{21} & \cdots & Y_{2n} & Y_{2(n+1)} & \cdots & Y_{2(n+m)} \\ \vdots & \ddots & \vdots & \vdots & \ddots & \vdots \\ Y_{n1} & \cdots & Y_{nn} & Y_{n(n+1)} & \cdots & Y_{n(n+m)} \\ \hline Y_{(n+1)1} & \cdots & Y_{(n+1)n} & Y_{(n+1)(n+1)} & \cdots & Y_{(n+1)(n+m)} \\ \vdots & \ddots & \vdots & \vdots & \ddots & \vdots \\ Y_{(n+m)1} & \cdots & Y_{(n+m)n} & Y_{(n+m)(n+1)} & \cdots & Y_{(n+m)(n+m)} \end{array}\right] \begin{bmatrix} V_1 \\ V_2 \\ \vdots \\ V_n \\ \hline V_{n+1} \\ \vdots \\ V_{n+m} \end{bmatrix} \quad (B1)$$

where $I_{bus}$ is the vector of the injected bus currents; $V_{bus}$ is the vector of bus voltages measured from the reference node (terminal bus of generator one for this document); $Y_{bus}$ is the bus admittance matrix: $Y_{ii}$ (diagonal element) is the sum of admittances connected to bus $i$ and $Y_{ij}$ (off-diagonal element) equals the negative of the admittance between buses $i$ and $j$.

Rewrite $(B1)$ into

$$\begin{bmatrix} 0 \\ I_m \end{bmatrix} = \begin{bmatrix} Y_{nn} & Y_{nm} \\ Y^t_{nm} & Y_{mm} \end{bmatrix} \begin{bmatrix} V_n \\ V_{t,m} \end{bmatrix} \qquad (B2)$$

$$I_m = [Y_{mm} - Y^t_{nm} Y^{-1}_{nn} Y_{nm}] V_{t,m} \qquad (B3)$$

Thus, all nodes other than the generator terminal nodes are eliminated. The admittance matrix of the reduced network to the terminal bus of generator is defined as

$$Y_t = Y_{mm} - Y^t_{nm} Y^{-1}_{nn} Y_{nm} \qquad (B4)$$

Note that the matrix $Y_r$ used in $(A1)$ is defined based on $(B4)$ and they, i.e. $Y_r$ and $Y_t$, convey the same information. Denote $Y_{t,ij}=G_{ij}+jB_{ij}$, which is the element of $Y_t$ in row $i$ and column $j$.

Then, $Y_r$ is defined as

$$Y_r = \begin{pmatrix} \begin{bmatrix} G_{11} & -B_{11} \\ B_{11} & G_{11} \end{bmatrix} & \cdots & \begin{bmatrix} G_{1m} & -B_{1m} \\ B_{1m} & G_{1m} \end{bmatrix} \\ \vdots & \ddots & \vdots \\ \begin{bmatrix} G_{m1} & -B_{m1} \\ B_{m1} & G_{m1} \end{bmatrix} & \cdots & \begin{bmatrix} G_{mm} & -B_{mm} \\ B_{mm} & G_{mm} \end{bmatrix} \end{pmatrix} \qquad (B5)$$

Note that $Y_r$ is a $(2m)\times(2m)$ matrix while $Y_t$ is $m\times m$.

## VII. APPENDIX C

This Appendix will show that under each of the following two conditions, the matrix $Y$ used in (2) and (4) will become a constant matrix: (i) for classical generator model; (ii) for fourth-order generator model with $X'_d=X'_q$.

For both classical generator model and fourth-order generator model, $(A4)$ becomes:

$$Z_k = \begin{pmatrix} R_{ak} & -X'_{dk} \\ X'_{dk} & R_{ak} \end{pmatrix} \qquad (C1)$$

So we have:

$$\begin{aligned} &Z_k \Delta_k = \Delta_k Z_k \\ &= \begin{pmatrix} R_{ak}\sin\delta_k - X'_{dk}\cos\delta_k & -X'_{dk}\sin\delta_k - R_{ak}\cos\delta_k \\ X'_{dk}\sin\delta_k + R_{ak}\cos\delta_k & R_{ak}\sin\delta_k - X'_{dk}\cos\delta_k \end{pmatrix} \end{aligned} \qquad (C2)$$

Then from $(A2)$, we have:

$$\begin{aligned} Y &= \left(T_1 Y_r^{-1} + T_2 T_1\right)^{-1} T_1 = \left(T_1 Y_r^{-1} + T_1 T_2\right)^{-1} T_1 \\ &= \left(T_1\left(Y_r^{-1} + T_2\right)\right)^{-1} T_1 \\ &= \left(\left(Y_r^{-1} + T_2\right)^{-1} T_1^{-1}\right) T_1 \\ &= \left(Y_r^{-1} + T_2\right)^{-1} \end{aligned} \qquad (C3)$$

where the product of $T_1$ and $T_2$ commute. Since $Y_r$ and $T_2$ are both constant matrices, $Y$ is constant.

## VIII. APPENDIX D

As an example, the pre-fault, fault-on and post-fault $Y$ matrices are shown below for a three-phase fault occurring near bus 7 at the end of line 5-7 and cleared in five cycles (0.083s) by opening line 5-7.

$$Y^{on} = \begin{bmatrix} 0.657 & 3.816 & 0 & 0 & 0.070 & -0.631 \\ -3.816 & 0.657 & 0 & 0 & 0.631 & 0.070 \\ 0 & 0 & 0 & 5.486 & 0 & 0 \\ 0 & 0 & -5.486 & 0 & 0 & 0 \\ 0.070 & -0.631 & 0 & 0 & 0.174 & 2.796 \\ 0.631 & 0.070 & 0 & 0 & -2.796 & 0.174 \end{bmatrix} \qquad (D1)$$

$$Y^{after} = \begin{bmatrix} 1.181 & 2.229 & 0.138 & -0.726 & 0.191 & -1.079 \\ -2.229 & 1.181 & 0.726 & 0.138 & 1.079 & 0.191 \\ 0.138 & -0.726 & 0.389 & 1.953 & 0.199 & -1.229 \\ 0.726 & 0.138 & -1.953 & 0.389 & 1.229 & 0.199 \\ 0.191 & -1.079 & 0.199 & -1.229 & 0.273 & 2.342 \\ 1.079 & 0.191 & 0.174 & 0.199 & -2.342 & 0.273 \end{bmatrix} \qquad (D2)$$